\documentclass{aa}  
\usepackage{graphicx}
\usepackage{xspace}
\usepackage{tabularx}
\usepackage{booktabs}
\usepackage{subfig}
\usepackage{txfonts}
\newcommand{\hh}{$\rm H_2$\xspace}
\newcommand{\kms}{$\rm km \, s^{-1}$\xspace}
\newcommand{\co}{$\rm C^\mathrm{18}O$\xspace}
\newcommand{\dco}{$\rm DCO^+$\xspace}
\newcommand{\cm}{$\rm \mathrm{cm}^{-2}$\xspace}

\begin{document}

   \title{The kinematics of the magnetized protostellar core IRAS15398-3359}

   \author{Farideh. S. Tabatabaei
          \inst{1}
          \and
          Elena Redaelli\inst{1}
           \and
          Paola Caselli\inst{1}
           \and
          Felipe O. Alves\inst{1}
          }

   \institute{Centre for Astrochemical Studies, Max-Planck-Institut für extraterrestrische Physik, Gießenbachstraße1, 85749 Garching bei München, Germany
             }

   \date{Received XXX; accepted XXX}


  \abstract
 {Observations of protostellar envelopes are essential in order to understand better the process of gravitational collapse toward star and planet formation. From a theoretical perspective, magnetic fields are considered an important factor during the early stages of star formation, especially during the main accretion phase.}
   {Our aim is to study the relation between kinematics and magnetic fields at a very early stage of the star formation process by using data from the Atacama Pathfinder EXperiment (APEX) single-dish antenna with an angular resolution of 28".}
   {We observed the two molecular lines \co (2-1) and \dco(3-2) toward the Class 0 young stellar object IRAS15398-3359. We implemented a multi-component Gaussian fitting on the molecular data to study the kinematics. In addition, we used previous polarization observations on this source to predict the influence of the magnetic field on the core.}
   {The velocity gradient along the central object can be explained as an ongoing outflow motion. We report the flowing of material from the filament toward the central object, and   the merging of two velocity components in the \co (2-1) emission around the protostar position, probably due to the merging of filamentary clouds. Our analysis shows that the large-scale magnetic field line observed previously is preferentially aligned to the rotation axis of the core.}
   {}


   \maketitle
%

\section{Introduction}

Observing protostellar envelopes around very young protostars is fundamental to gaining a better understanding of the progression of the collapse of protostellar cores toward planetary systems. Class 0 objects are known to represent the youngest stage of protostellar evolution (\citealt{andre93}, \citealt{andre2000}). Most of their mass is contained in a dense envelope that accretes to the central protostar during the main accretion phase (\citealt{maury2011}, \citealt{evans2009}). Protostars are deeply embedded in their parent cores, which may cause interactions between protostellar outflows and surrounding gas leading to complex morphologies. Detailing these structures will allow us to learn about the dynamics of protostellar evolution at an early stage. In light of these circumstances, it becomes essential to investigate in detail the earliest stages of star formation for specific sources.

The study of the extent and contribution of magnetic fields in star formation and the competition between magnetic and turbulent forces is still a highly debated topic in modern astronomy (e.g., \citealt{mclow}; \citealt{mckee}; \citealt{crutcher}).  However, in the star formation process, especially during the early stages, magnetic fields (B) are expected to play a crucial role, providing a source of non-thermal pressure against the gravitational pull \citep{mckee}.
In light of the fact that interstellar gases are often mildly ionized \citep{caselli1998}, the matter is likely to be coupled with the magnetic field lines at envelope scales. Due to gravity, magnetic lines bend inward, thus producing an hourglass shape, and in low-mass star-forming regions this effect is not detected frequently (detected in 30 percent of young stellar objects in polarization; \citealt{hull}, \citealt{pattle2022}).

IRAS15398-3359 (hereafter IRAS15398) is a low-mass Class 0 protostar at a distance of 156 pc \citep{dzib2018}, embedded in the Lupus I molecular cloud, $\alpha_{2000} = 15^\mathrm{h}43^\mathrm{m}02^\mathrm{s}.2,  \delta_{2000} = -34^\circ09^{'}07.7^{''}$. It has a bolometric temperature of 44 K \citep{jorgensen2013}. The protostellar mass is 0.007$^{+0.004}_{-0.003} \mathrm{M}_\odot$ \citep{okoda2018A}. 
Lupus I is the least evolved component of the Lupus complex \citep{rygl2013}, and optical polarization studies have demonstrated that Lupus I is threaded by a very ordered magnetic field that is perpendicular to its filamentary extension \citep{franco}. Therefore, it is an ideal place to study the kinematics of the early stages of low-mass star formation and the connection between the source kinematics and the strong large-scale magnetic field. By observing its CO emission line with the single-dish and interferometric observation, a molecular outflow was detected from this source (\citealt{tachihara1996}; \citealt{bjerkeli}; \citealt{van-kempen}). The core is embedded in a less dense $(N(H_2) \sim  10^{22} $ \cm) filamentary structure, which extends toward the  northwest.

Based on magnetic field studies in protostellar core simulation analysis, more magnetized cores show strong alignments of the outflow axis with the magnetic field orientation, whereas less magnetized cores display more random alignment \citep{lee2017}. Observational results present a mixture of cases. \cite{galametz2018} used a sample of 12 low-mass Class 0 protostars to investigate the submillimeter polarized emission at scales of $\sim$ 600 - 5000 au, and demonstrated a relation between the field morphology, the core rotational energy, and the multiplicity of the protostellar system. According to that paper's analysis, the envelope scale magnetic field  tends to be either aligned or perpendicular to the outflow direction, but for single sources the magnetic field is aligned along the outflow direction. \cite{yen2021} studied 62 low-mass Class 0 and I protostars in nearby (<450 pc) star-forming regions with the orientations of the magnetic fields on 0.05–0.5 pc scales. They suggest that the outflows are likely to be misaligned with B-fields by 50 degrees in 3D space. While \cite{hull2019} used Atacama Large Millimeter/submillimeter Array (ALMA) observations with spatial resolutions of up to $\sim$ 100 au, and conclude that magnetic fields and outflows are randomly aligned in low-mass protostellar cores. The discrepancy between simulations and observations can be due to the limitations of the simulation setup. As an example, \cite{lee2017} applied an ideal magnetohydrodynamic (MHD) simulation when in reality non-ideal MHD effect might be important \citep{wurster2021}.

\cite{redaelli2019} used polarimetric observations of the dust thermal emission at 1.4THz obtained with the Stratospheric Observatory for Infrared Astronomy (SOFIA) telescope to investigate the magnetic field properties at the core scales toward IRAS15398. The authors found a uniform magnetic field consistent with the large-scale field derived from optical observations \citep{franco}. They suggested the core experienced a magnetically driven collapse and the core inherited the B-field morphology from the parental cloud during its evolution. The field lines pinch inward toward the central object, leading to the characteristic hourglass shape that is predicted by models of magnetically driven collapse. 
They showed that the mean direction of the magnetic field is aligned with the large-scale B-field and with the direction of outflow. Their prediction for a  magnetic field strength of B = 78 $\mu$G  is expected to be  accurate within a factor of two. They calculated the mass-to-flux ratio, $\lambda = 0.95$, which means that the core is in a state of transition between supercritical and subcritical states.

In this paper we present new observational data that allow us to study the gas kinematics, which we compare   to the magnetic field direction. The aim is to assess the importance of magnetic fields in the dynamical evolution of low-mass star-forming regions. The outline of the paper is as follows. The observations and results are described in detail in Sects. 2 and 3. In Sect. 4 we analyze the observed line profiles of \co (2-1) and \dco (3-2) emission lines. The results are summarized in Sect. 5.

\section{Observations}

\subsection{APEX}

IRAS15398 was observed using the Atacama Pathfinder EXperiment (APEX) single-dish antenna located at Llano de Chajnantor in the Atacama desert of Chile on 2019 September 14, 16, 17, 21, and 23. We used the PI230 receiver coupled with the FFTS4G backend in the on-the-fly mode. We used the highest spectral resolution that the FFTS4G could provide, 62.5 kHz ($\approx$ 0.08 \kms at the frequency of the \dco(3-2) line). The data were reduced to a pixel size of 9 arcsec. The broad bandwidth of the PI230 receiver can be set up to observe simultaneously \co (2-1) and the \dco(3-2) transitions at 219.560 GHz and 216.113 GHz, respectively, the target lines of this research.
The angular resolution at these frequencies is $\sim$ 28 arcsec, corresponding to 0.02 pc at the source distance 156 pc \citep{dzib2018}.
The data reduction was performed based on the standard procedure of the CLASS software, GILDAS.\footnotetext[1]{\url{https://www.iram.fr/IRAMFR/GILDAS/}}\footnotemark[1]
The antenna temperature, $T_A$, was converted to main-beam brightness temperature using the forward efficiency ($\eta_\mathrm{fw}$) and main beam efficiency, $T_\mathrm{mb} = T_\mathrm{A} \frac{\eta_\mathrm{fw}}{\eta_\mathrm{mb}}$, 
given a main beam efficiency $\eta_\mathrm{mb}$ = 0.8.\footnotetext[2]{\url{www.apex-telescope.org/telescope/efficiency/}}\footnotemark[2]

\subsection{Herschel}
We used archival data from the Gold Belt Survey, obtained with the \textit{Herschel} space telescope to obtain the gas column density and the dust temperature (\citealt{andre2010}, \citealt{rygl2013}, \citealt{benedettini2018}) . The N(\hh) column density map has a resolution of $\sim 38$ arcsec.

\begin{figure}
    \centering
    \subfloat{{\includegraphics[width=8.5cm]{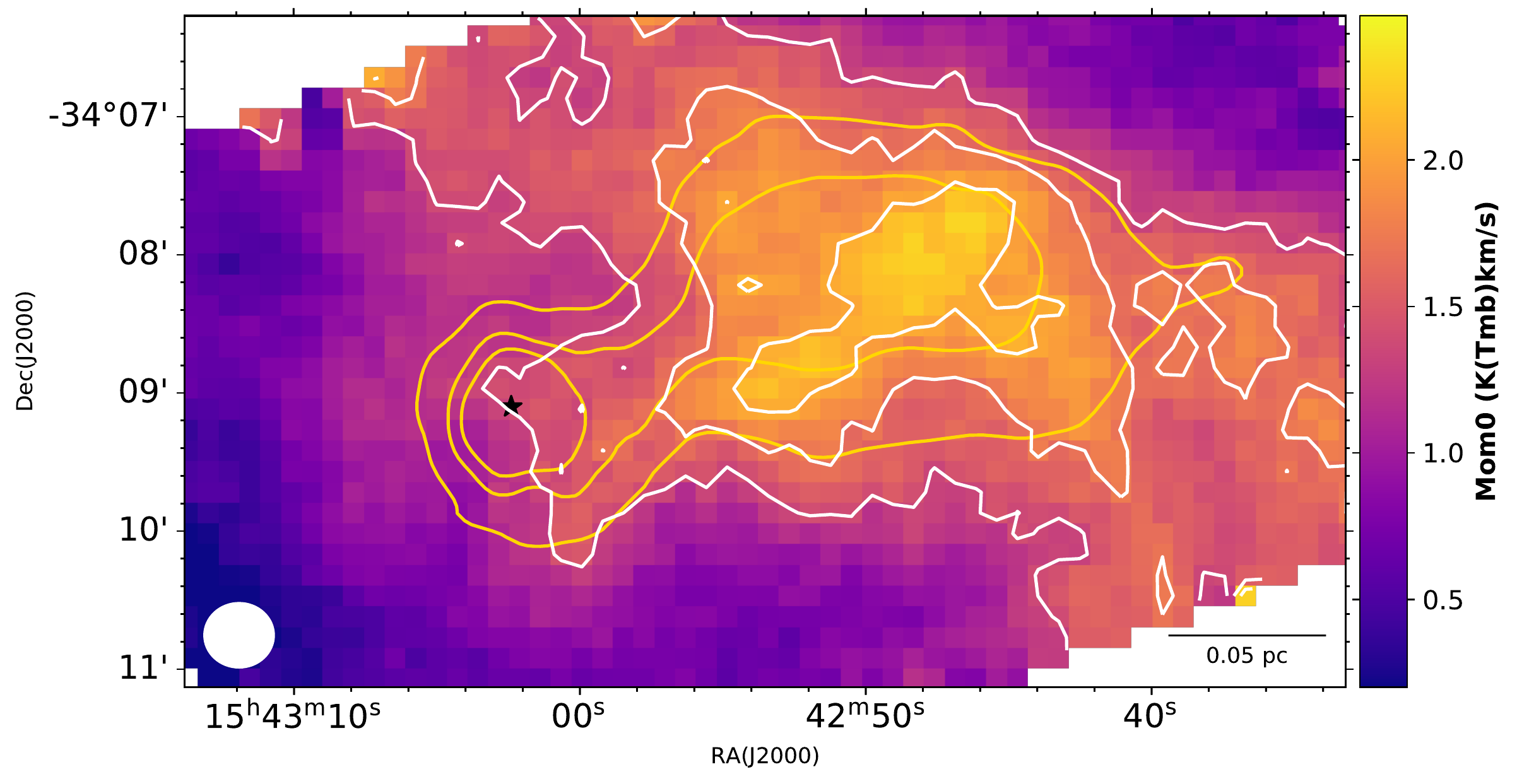} }}
    \qquad
    \subfloat{{\includegraphics[width=8.5cm]{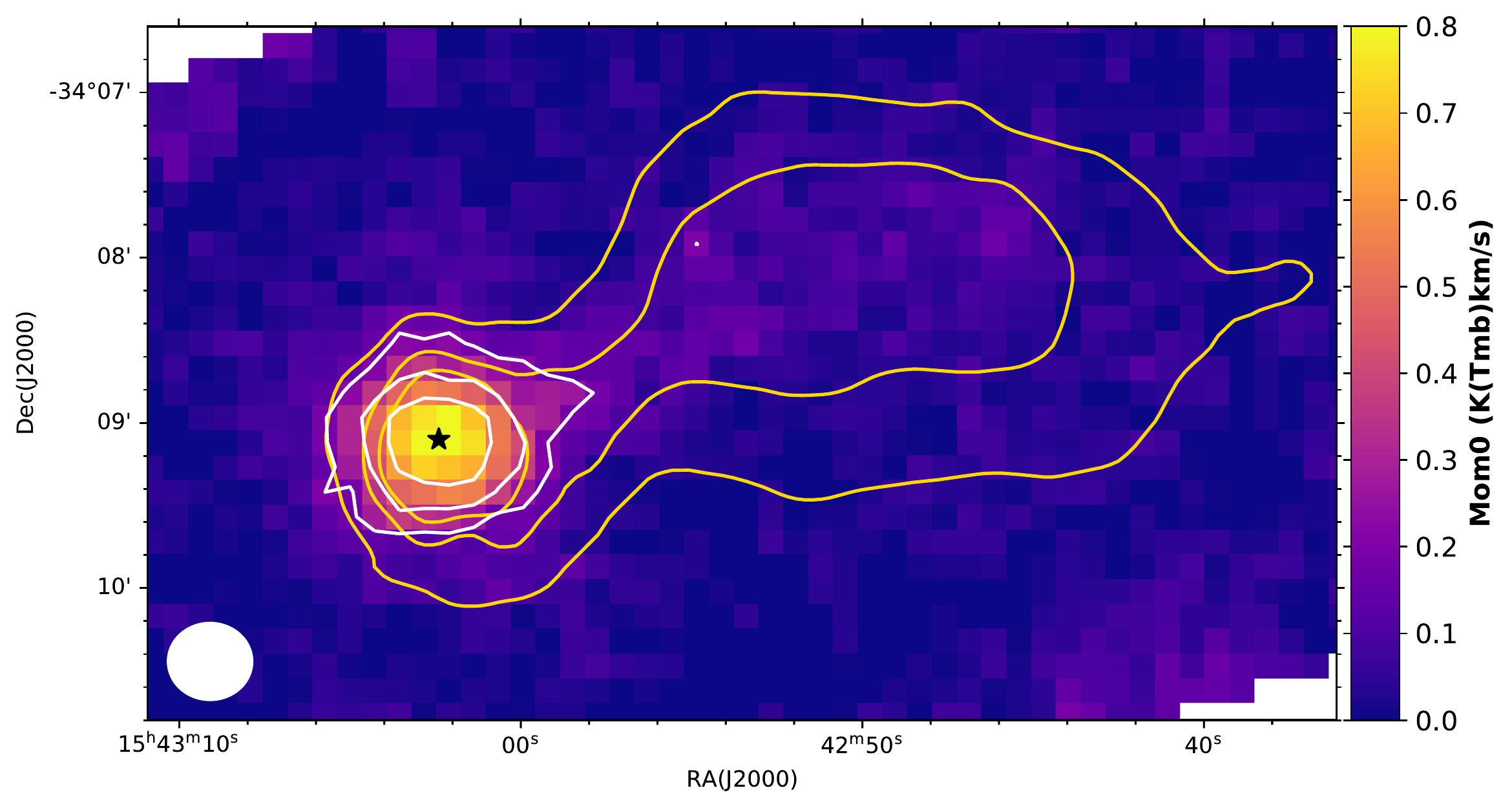} }}
    \caption{Integrated intensity of \co(1-2) (top) and \dco(3-2) (bottom) toward IRAS15398. The white contour levels are 10, 20, and 30 times of mean rms value for the \dco line and 40, 50, and 60 times of mean rms value for the \co (2-1) line. The beam size is shown in the bottom left corner. The yellow contours in both images show \hh column density (levels:[1.0, 1.5, 2.0] $10^{22}$ $\mathrm{cm}^{-2}$). The black star gives the position of the protostar.}
    \label{moment}
\end{figure}

\section{Result}

\begin{figure*}
                \begin{center}
                \includegraphics[width=0.98\textwidth]{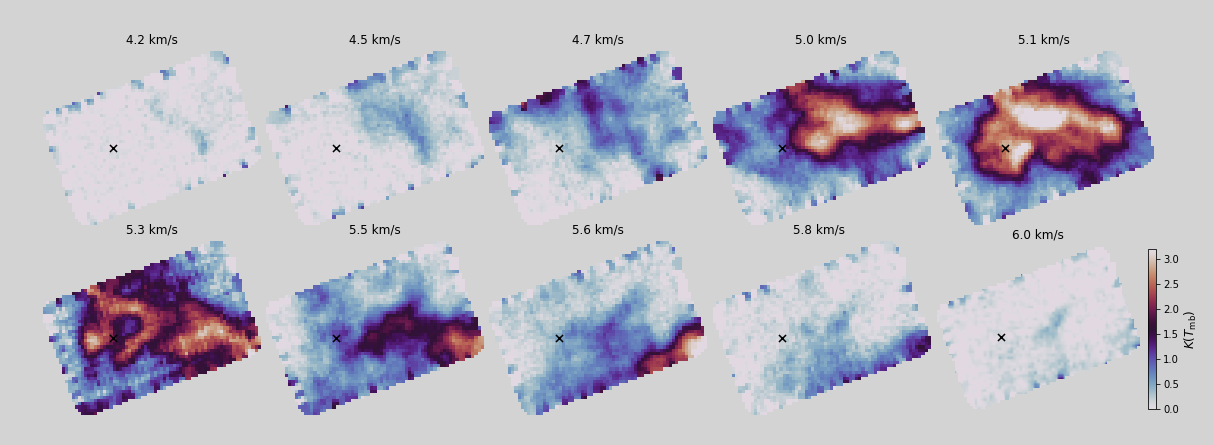}
                \caption{Channel maps of \co (2-1) emission. The black cross indicates the position of the protostar core. The velocity of each channel is shown above each panel.}
                \label{channelc18o}
                \end{center}
        \end{figure*}

Figure \ref{moment} presents the moment 0 (integrated intensity) maps of \co (2-1) and \dco (3-2) overlaid with the contours of \hh column density. The mean rms in $T_\mathrm{mb}$ scale is 0.1 K and 0.08 K for  \co (2-1) and \dco (3-2), respectively. On the basis of emission-free channels only, we derived the mean rms per channel for each transition. This rms is used to associate uncertainties at each pixel to the integrated intensity, the mean value for \co(2-1) and \dco(3-2) being 0.03 K \kms and 0.02 K \kms, respectively. These lines are optically thin and do not present crowded hyperfine structures. In order to confirm this hypothesis, we evaluate the opacity as

\begin{equation}
    \tau = - \ln \left[ 1- \frac {T_\mathrm{MB}} {J_\nu(T_\mathrm{ex}) - J_\nu(T_\mathrm{bg})} \right],
\end{equation}
where $T_\mathrm{ex} = 12$ K is the excitation temperature (obtained from the \emph{Herschel} dust temperature map, see also Sect. 4.1), $T_\mathrm{MB} = 3.5$\,K is the peak main beam temperature,   the function J$_\mathrm{\nu}$ is the equivalent Rayleigh--Jeans temperature, and $T_\mathrm{bg} = 2.73$ K is the cosmic background temperature. Therefore, we obtain $\tau = 0.66$. For the \dco (3-2) line we obtain $\tau = 0.44$ using $T_\mathrm{MB} = 1.0$ and $T_\mathrm{ex} = 7$ K.
Through a combination of excitation and abundance, distinctive species give complementary information on gas conditions. Due to its relatively high abundance and low critical density 
$n_\mathrm{cr}$(\co(2-1)) $\sim 10^4 \mathrm{cm}^\mathrm{-3}$,
computed with numbers in the LAMDA database,\footnotetext[3]{\url{https://home.strw.leidenuniv.nl/~moldata/}}\footnotemark[3] \co (2-1) is a sensitive tracer of relatively low-density material in the cloud, which traces the more extended gas in the filamentary structure. 
The \dco(3-2) molecule, on the other hand, presents a higher critical density $n_\mathrm{cr}(\rm DCO^+(3-2)) \sim 10^6 \mathrm{cm}^\mathrm{-3}$, which makes it more selective of dense gas closer to the central protostar. 
\dco (3-2)  emission is also known as a remarkably sensitive tracer for gas properties during the early stages of protostellar evolution  (e.g., \citealt{gerner2015}).  In the location of the protostar, we see a  decrease in the \co (2-1) integrated intensity, suggesting that the molecule is partially depleted onto the dust grains (\citealt{caselli1999}, \citealt{bacmann2002}).
Figure \ref{channelc18o} shows the channel maps of the \co (2-1) line. The figure presents the signal emission at velocity intervals of $\approx$ 0.2 \kms.


 \section{Analysis}

 \subsection{Column density maps}

To calculate the column density map of these two molecules we used the same procedure as used in \cite{redaelli-density} and \cite{caselli2002a} for an optically thin transition. The \dco (3-2) and \co (2-1) lines are both optically thin, as we show in the previous section. The expression of the total column density derived by an optically thin transition is given by

\begin{equation}
N_\mathrm{col} = \frac{8 \pi W \nu^3}{c^3  A_\mathrm{ul} } \frac{Q(T_\mathrm{ex})}{J_\mathrm{\nu}(T_\mathrm{ex}) - J_\mathrm{\nu}(T_\mathrm{bg})} \frac{e^\mathrm{\frac{E_\mathrm{u}}{k_\mathrm{B} T_\mathrm{ex}}} }{g_\mathrm{u}(e^\mathrm{\frac{ h \nu}{k_\mathrm{B} T_\mathrm{ex}} }-1) },
\end{equation}where $T_\mathrm{ex}$ is the excitation temperature, the function J$_\mathrm{\nu}$ is the equivalent Rayleigh--Jeans temperature, $T_\mathrm{bg}$ = 2.73 K is cosmic background temperature, $E_\mathrm{u}$ is the upper state energy, $g_\mathrm{u}$ is the degeneracy, $A_\mathrm{ul}$ is the Einstein coefficient, $Q$ is the partition function, $\nu $ is the line frequency, $h$ is the Planck constant,  k$_\mathrm{B}$ is the Boltzmann constant (see Table 1 for details),\footnotetext[4]{\url{https://cdms.astro.uni-koeln.de/cgi-bin/cdmssearch}}\footnotemark[4] and $W$ is the integrated intensity of the line. Since the \dco (3-2) transition shows only one velocity component, we use the result of the Gaussian fit to compute the integrated intensity of this line (see Sect. 4.2 for more details) by calculating the area under the Gaussian profile. The \co (2-1) emission, instead, shows signs of multiple velocity components along the line of sight. We therefore compute the integrated the intensity from the data cube, integrating emission over the velocity range [4 - 6.5]\kms, which contains the whole line profile.

\begin{table}[ht]

\caption{Spectroscopic parameters used to derive the molecular column density.} 
\centering 
\renewcommand{\arraystretch}{1.1}

\begin{tabular}{c c c c c c} 
\hline \hline
\toprule
Transition & $\nu$ & $g_\mathrm{u}$ & $E_\mathrm{u} /10^{-22}$ & $A_\mathrm{ul}/10^{-3}$ & Q(7$^a$) \\  
          & (GHz)   &   & (J)   & (s$^{-1}$) &    \\[0.7ex]

\midrule

\co(2-1)  & 219.56  & 5 & 2.18 & 6.01 10$^{-4}$ & -    \\ 
\dco(3-2) & 216.112 & 7 & 2.86 & 7.65 & 4.40$^b$  \\ [1ex]  

\hline \\ 
\end{tabular}
\label{Parameters} 
\footnotesize \textbf{Note}: All the data are   from the Cologne Database for Molecular Spectroscopy (CDMS) documentation.\footnotemark[4]
\footnotesize{$^a$ The excitation temperature of \co (2-1) is 7K. $^b$ Calculated via interpolation of the Partition function available at CDMS, for different temperatures.  }\\

\end{table}

We use the dust temperature map to approximate the excitation temperature for the \co (2-1) line, obtained from \emph{Herschel} data (\citealt{benedettini2018}, \citealt{rygl2013}) since we expect this line to be thermally excited. This assumption may induce some small errors as at the volume densities traced by the \co (2-1) line the gas and dust are not thermally coupled \citep{goldsmith2001}.
We therefore use the dust temperature as a proxy for the gas kinetic temperature. On the contrary, the \dco (3-2) line is the 3-->2 transition, which has a high critical density; therefore, the dust temperature is not a good approximation for excitation temperature because we expect the line to be subthermally excited, so we use excitation temperature equal to 7 K with a variation of 2 K for this line. The column density peak for the \dco and \hh is found at the protostar position. For the protostar position, the \co, \dco, and \hh column densities are $ (7.8\pm 0.1) \times 10^{14}$\cm , $ 1.4_{-0.5}^{+2.24} \times 10^{12}$\cm, and $ (4.2 \pm 1.8) \times 10^{22}$ \cm \citep{roy2014}, respectively. The column density for \hh is obtained based on the \emph{Herschel} map.

\begin{figure*}
                \begin{center}
                \includegraphics[angle=-90,width=0.80\textwidth]{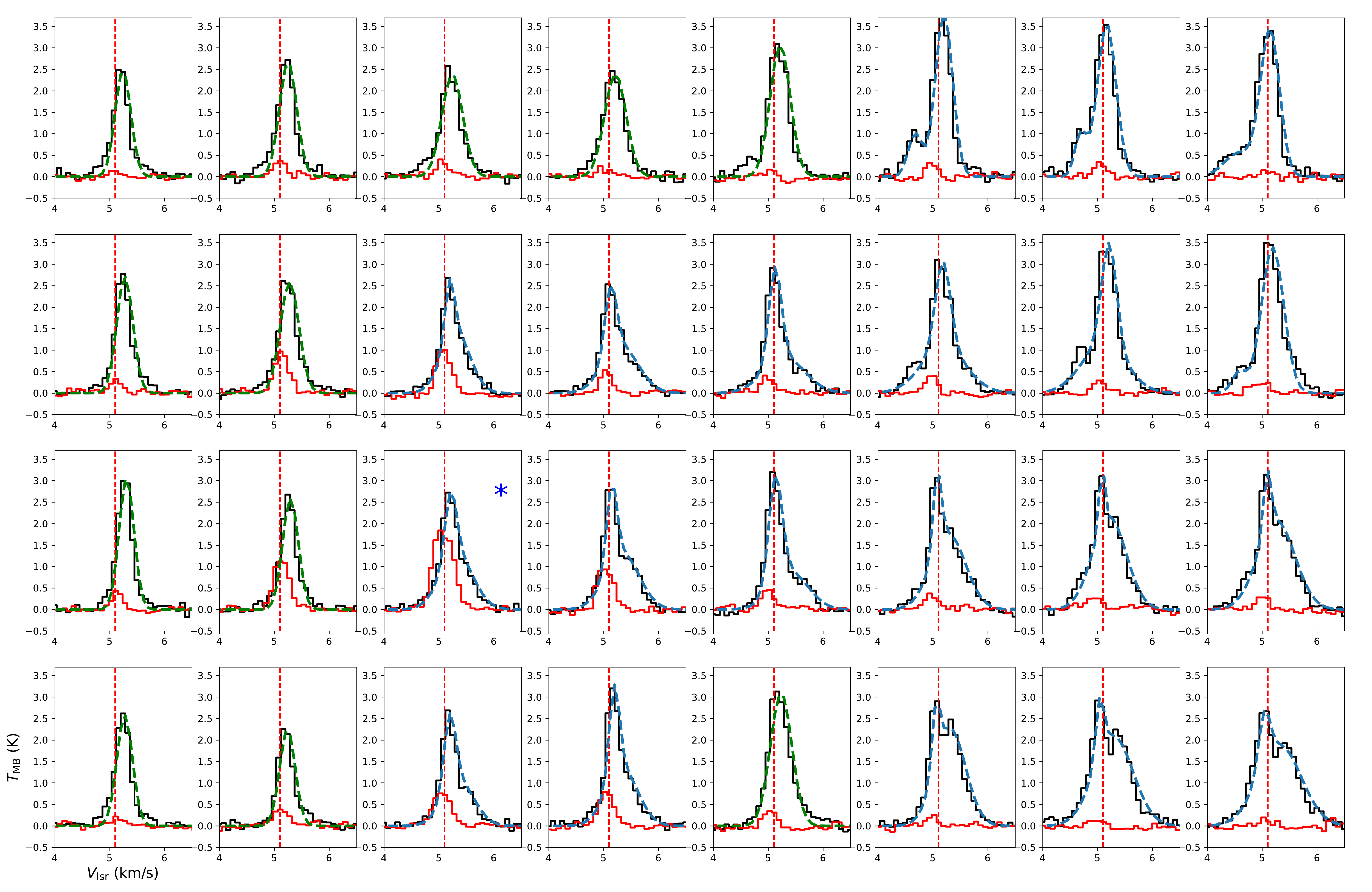}
                \caption{Spectrum grid of  \co (2-1) (black histogram) and \dco (3-2) (red histogram)   for 40 random positions around the core. Shown are the   two-Gaussian fits (blue curves) and one-Gaussian  fits (green curves) to the \co (2-1) line. The vertical line is V$_\mathrm{lsr}$=5.2 \kms. The blue star shows the spectrum for the position of the central protostar.}
                \label{channel}
                \end{center}
        \end{figure*} 


 \subsection{Spectral line analysis }
 
In order to derive the kinematics parameter maps (e.g., V$_{lsr}$, $\sigma_\mathrm{V}$), we perform a Gaussian fitting of the transitions using the \textsc{pyspeckit} package of python \citep{ginsburg2011}. For the \dco (3-2) data cube, we use a single-Gaussian component fit. The initial guesses are then 2.5 K, 5.2 \kms, and  0.2 \kms for the amplitude, velocity dispersion, and width, respectively. 

Instead, \co (2-1) presents more complex kinematics. It often shows two velocity components in its profiles. Since the line is optically thin, as shown in Sect. 3, we are confident that these are multiple velocity components and they are not due to self-absorption. In order to fit two Gaussian profiles on the \co (2-1) data we perform a simple S/N cut, we mask those pixels where $\frac{T^\mathrm{peak}_\mathrm{MB}}{rms}$ < 20, then fit one Gaussian profile to all the unmasked pixels (65\% of the pixels). Then we fit two Gaussians for those pixels that had a residual larger than $2 \times rms$ and those Gaussian fits with a width broader than 0.25 \kms in the previous step. By  visual inspection we find that lines broader than 0.25 \kms show profiles consistent with two velocity components on the line of sight. For the second time we check the Gaussian fit profiles and for those pixels that have residuals bigger than $2 \times rms$ we do the fitting one more time with different initial guesses (60\% of the pixels with S/N cut), which means  they always have residuals less than two times the rms. As a final step, if the error on the velocity dispersion or velocity and on the amplitude is larger than 1 \kms and 1 K, respectively, we remove the fit. By doing this, we remove fits with unreasonably large uncertainties, which indicates that they have been poorly fitted. Figure \ref{channel} shows the fit results overlapped with the data.

The grid of spectra of \co (2-1) and \dco (3-2) lines for 40 positions at 18 arcsec intervals from each other around the core is shown in Fig. \ref{channel} (for more details about the position of each spectrum, see Appendix A). The red histogram represents the \dco (3-2) spectrum and the black histogram is the \co (2-1) spectrum. The blue curves represent the fit when the code uses a two-Gaussian and the green curves are the fit when the code used  one Gaussian for the \co (2-1) line. The vertical line is $V_\mathrm{sys}$ = 5.2 \kms, which is the systematic velocity computed from \co (2-1) line which was reported in \cite{yen2017} ($V_\mathrm{sys} = 5.24 \pm 0.03$  \kms), which is consistent with the systematic velocity seen in our data. 

The \dco (3-2) spectra always show a single velocity component with a Gaussian profile.
This line traces only one velocity component that is associated with the high-density material. On the other hand, as already mentioned, \co (2-1) shows two velocity components in their profiles. The brightest component is the one close to the velocity of the \dco (3-2) line, and it hence arises from the high-density material. The fainter component of \co (2-1) can appear on the red or blue side of the stronger component depending on the location in the cloud. In total, there are three different velocity components, the main one at 5.2 km/s (in all panels on all \dco (3-2) spectra locations), a lower velocity component seen in the northwestern panels, and a higher velocity component seen in the southwestern panels.
In the first two rows of Fig. \ref{channel} the faint component appears on the blue side, and in the two bottom rows of the spectrum grid it appears on the blue side of the stronger component. 

In Fig. \ref{channel} we can see the line profiles change across the grid. The multiple velocity components of \co (2-1) along the line of sight appear to merge from east to west (moving toward the position of the protostar).
\co (2-1) is a lower density tracer than \dco (3-2); these fainter velocity components, seen only in the former tracer, are likely additional nearby low-density filamentary clouds along the line of sight. The core envelope then appears to be found in the correspondence of the merger of these structures. In the following subsections, we discuss the maps of the kinematics parameters for each tracer individually.


\subsection{ $\rm DCO^+$ (3-2) line}

\begin{figure}
    \centering
    \subfloat{\includegraphics[width=8.5cm]{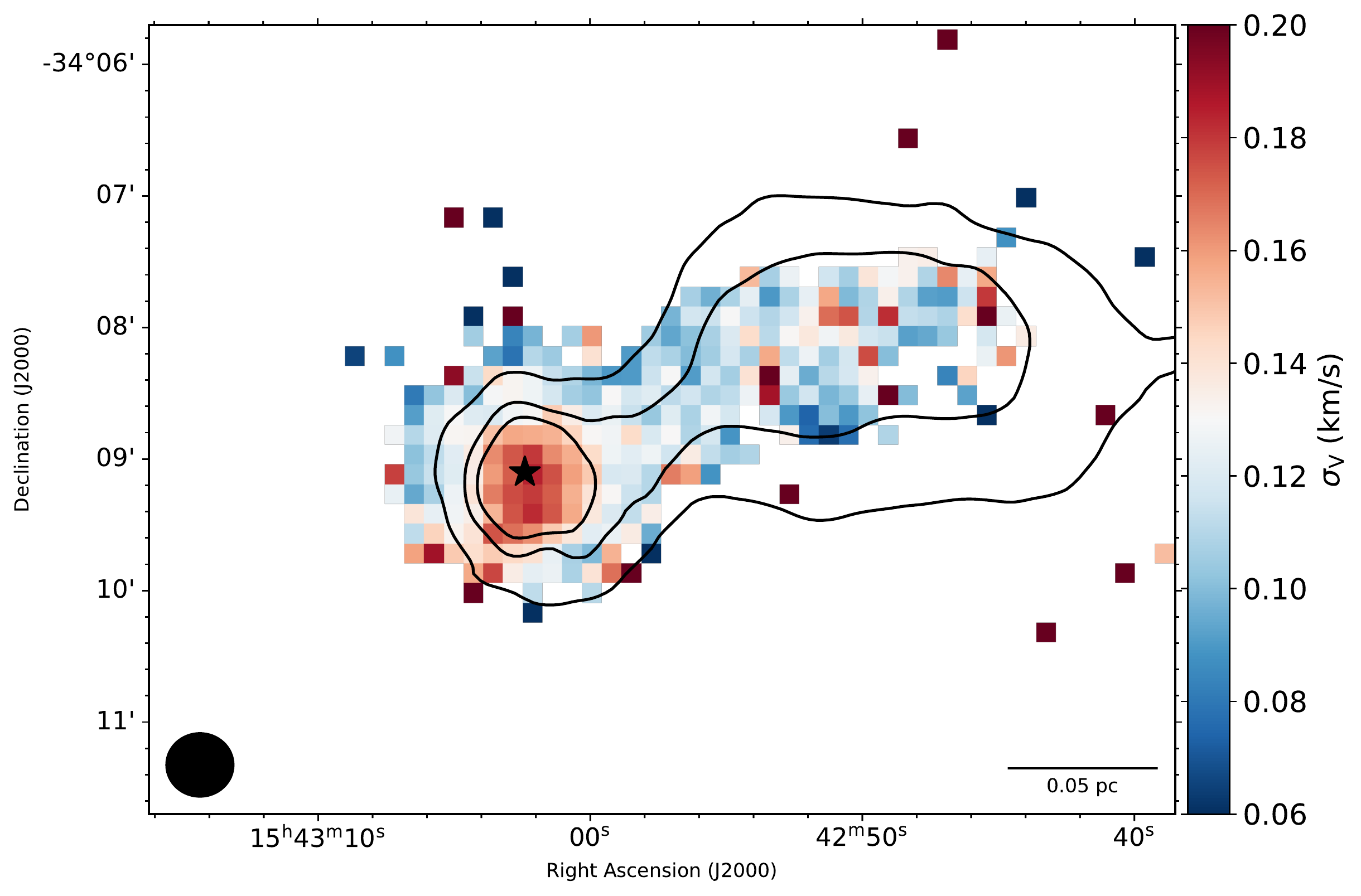} }
    \qquad
    \subfloat{\includegraphics[width=8.5cm]{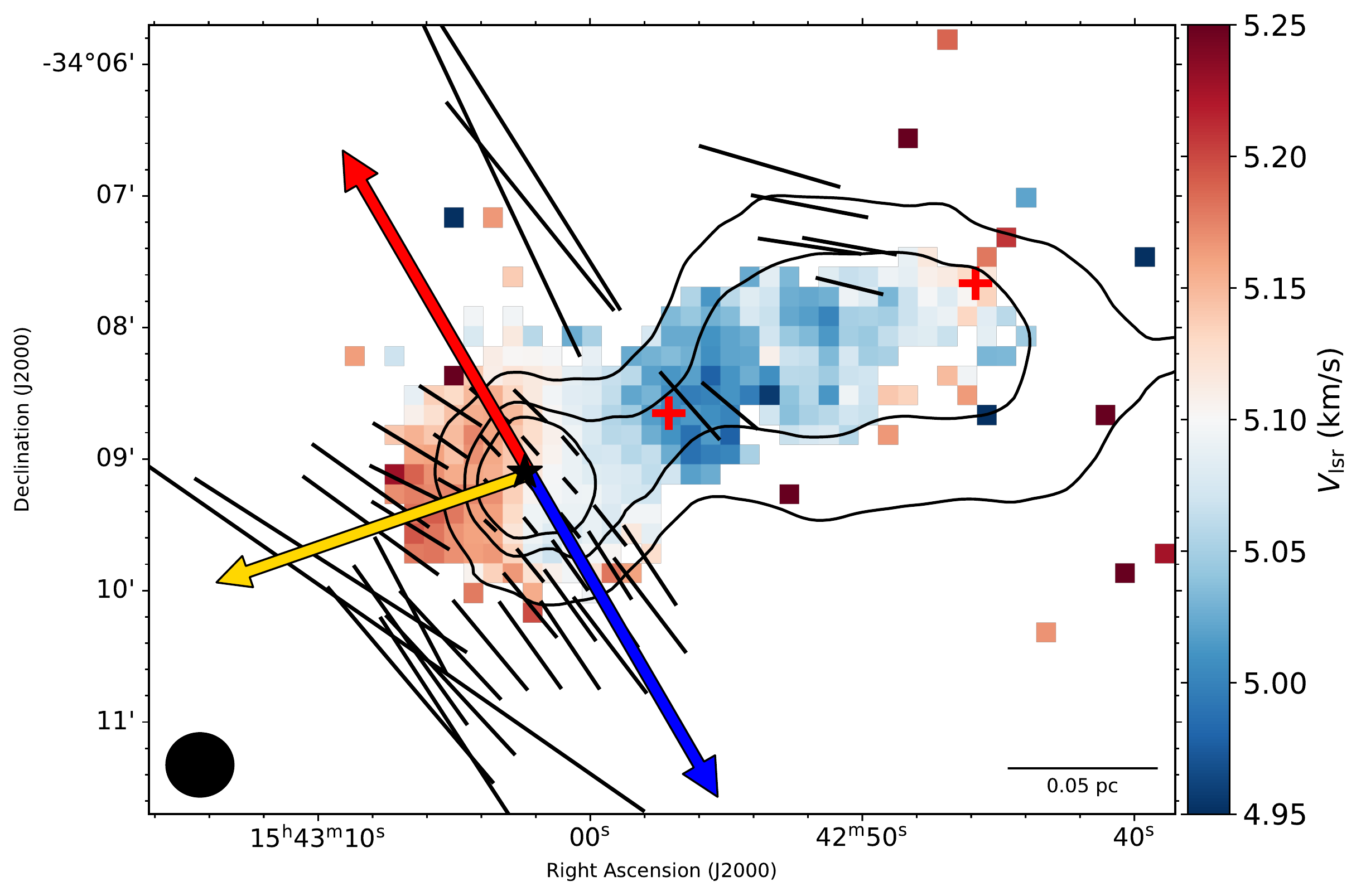} }
    \caption{Maps of kinematic parameters. Top: Gas velocity dispersion of the cloud traced using the \dco (3-2). Bottom: Centroid velocity map obtained fitting the observed \dco (3-2). The black vectors represent the polarization angles, tilted by 90 degrees to trace the magnetic field direction \citep{redaelli2019}. The blue and red arrows show the direction of the outflow ($PA=35^{\circ}$, from \citealt{bjerkeli}), and the yellow arrow presents the mean velocity gradient direction around the core. The star shows the position of the protostar. The contours represent N(\hh) column density levels,  as derived from \emph{Herschel} data: [1.0, 1.5, 2.0] $10^{22}$ \cm.}
    \label{velocity}
\end{figure}

 \begin{figure}
   \centering
   \includegraphics[width=\hsize]{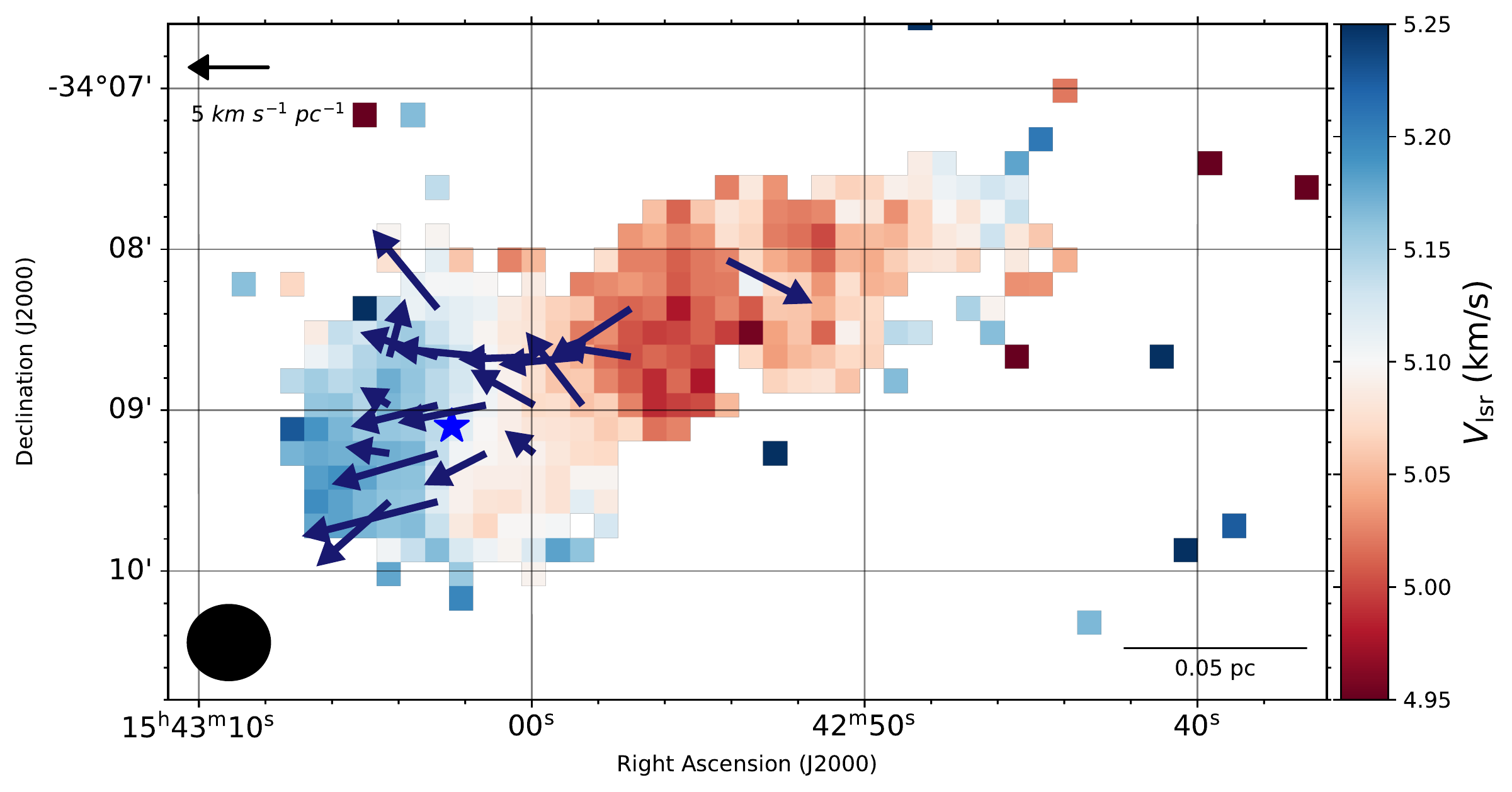}
      \caption{Centroid velocity map
of the \dco (3-2) line  overlaid with the gradient arrows (only vectors with $S/N>3$ are shown). The arrow length represents the relative vector magnitude of the gradient, according to the scale shown in the top left corner, and the direction of the arrows points to the steepest velocity field change. }
         \label{gradient_velocity}
   \end{figure}
%

 \begin{figure*}
    \centering

\includegraphics[width=0.47\textwidth]{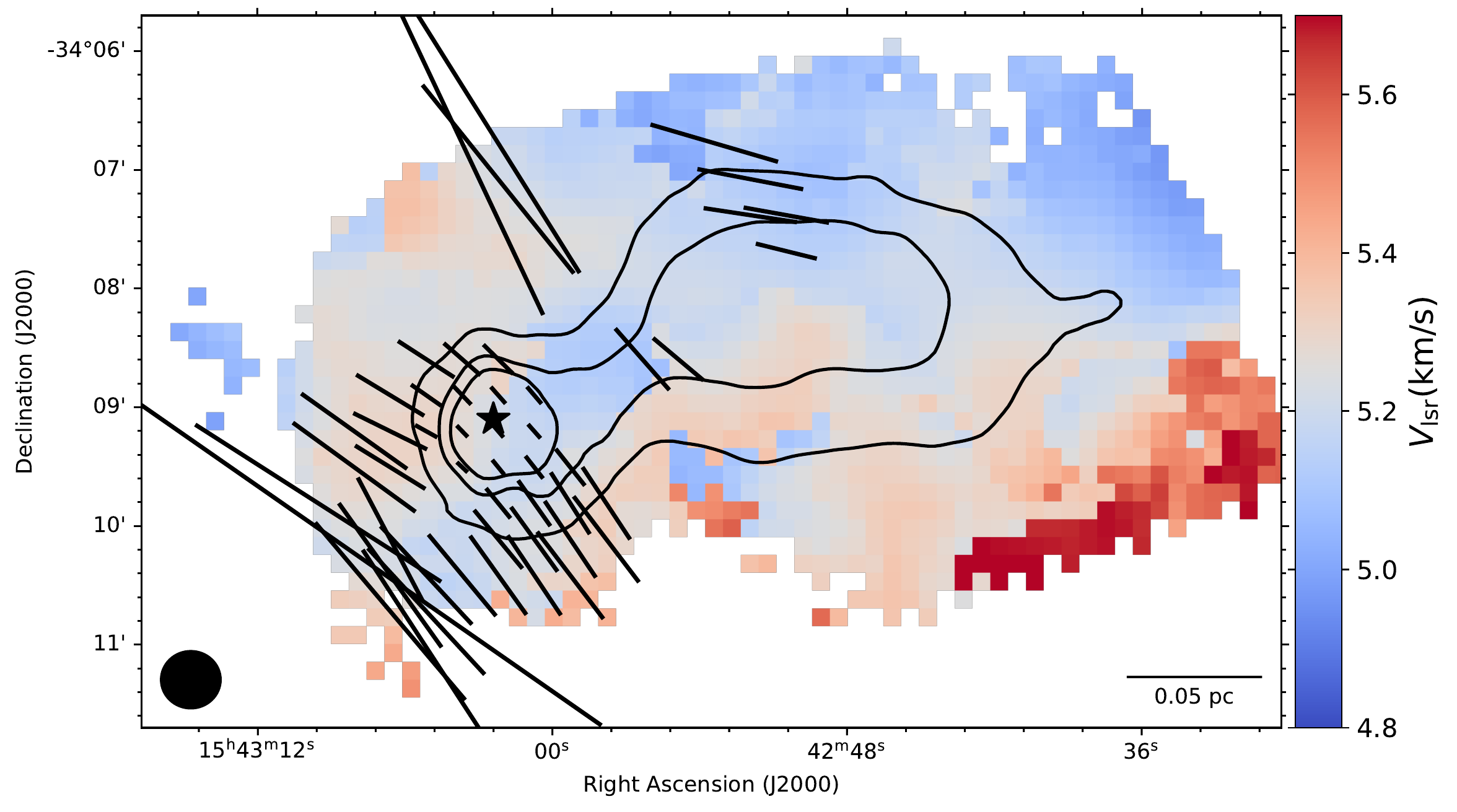}
\hfill
     \includegraphics[width=0.51\textwidth]{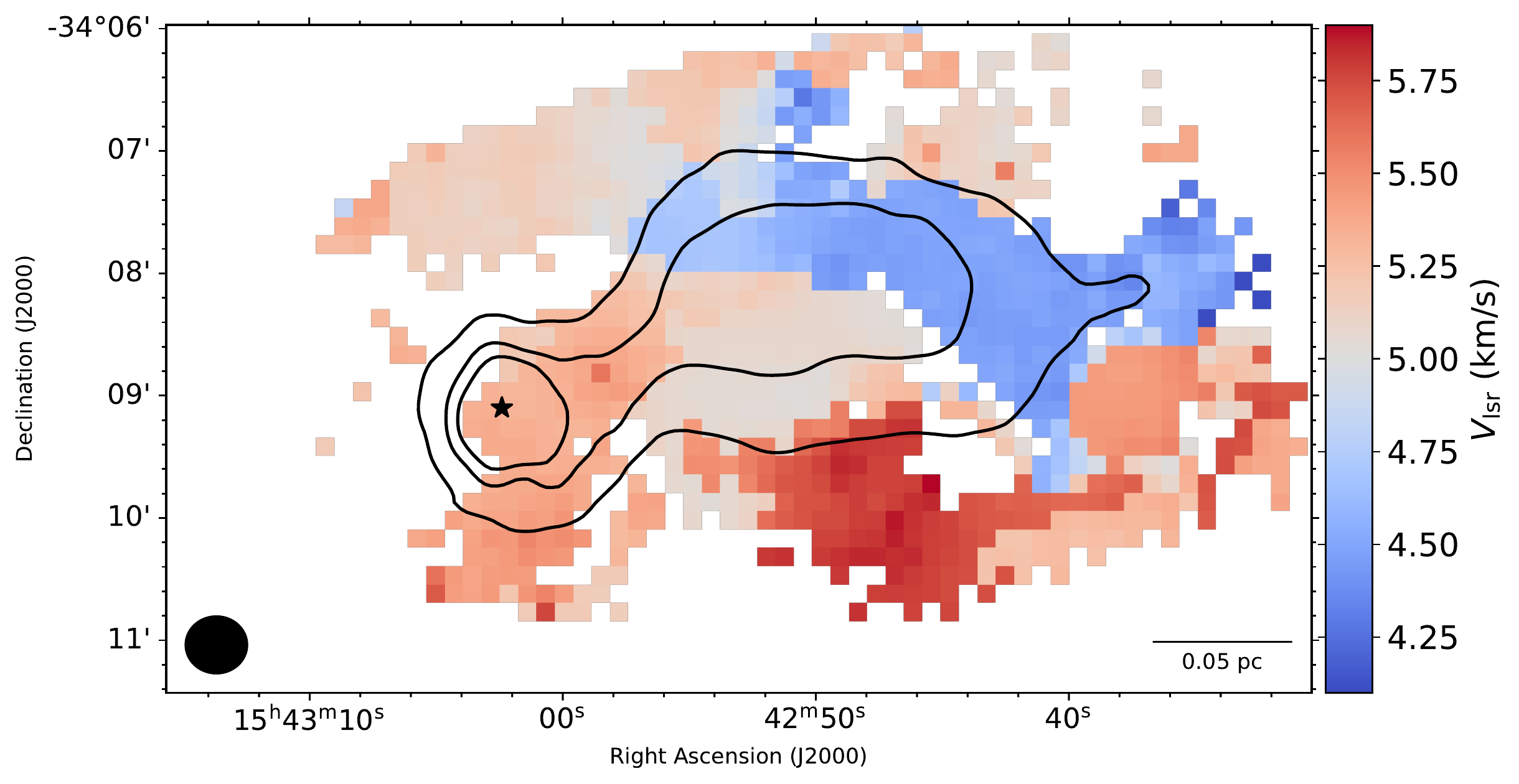}
\hfill    
        
     \caption{ Maps of kinematic parameters. Left panel: Bright component of centroid velocity map of \co (2-1). Right panel: Centroid velocity map only showing the weak component of the Gaussian fitting. The black vectors represent the polarization angles, tilted by 90 degrees. The star shows the position of the protostar. The contours represent N(H$_{2}$) column density levels, as derived from Herschel data: [1.0,1.5,2.0] $10^{22}$ \cm. }
     \label{C18O_V}
\end{figure*}
  

In Fig. \ref{velocity} (top panel) we show the velocity dispersion map, which presents a clear increase toward the center of the protostar envelope, starting with very narrow lines in the outer region (around the filament). The mean velocity dispersion derived from \dco (3-2) of the gas in the filament is in fact  $\langle \sigma_\mathrm{V} \rangle = 0.12 \pm 0.02$ \kms, but it becomes broader toward the center; for the positions where N(\hh) $> 3.2 \times 10^{22}$ \cm (inner contour in the Fig. \ref{velocity}), we derive $\langle \sigma_\mathrm{V} \rangle = 0.175 \pm 0.006$ \kms. This increase is linked to the protostellar activity, injecting turbulence, and  it could also be caused by the rotation of the core. 

We also observe oscillatory motions in the velocity field (visible in the \co data as well, see Fig. \ref{ppv_plot}, the green dots), which have been seen before in the large-scale velocity patterns. This velocity pattern is consistent with core-forming motions \citep{hacar2011}.
A small velocity gradient can be seen along the filament, which could be linked to the ongoing accretion material toward the central object. 

The filament extends over $\sim$ 0.1 \kms in velocity toward the protostar from 5.1 \kms in the west side to 5.0 \kms at the eastern edge. These positions are shown with red plus signs in Fig. \ref{velocity}. We determine a conservative estimate of the length of the filament of 0.11 pc.  This value is computed based on the \hh column density from \emph{Herschel} data, an area with a higher value than $1.5 \times 10^{22}$ \cm. In this border we  find almost all of  the filament (second contour in Fig. \ref{velocity}). 
Thereby the velocity gradient is $ \nabla V = \Delta V / \Delta R = 0.91 \pm 0.23$  $\mathrm{km} \mathrm{s}^{-1} \mathrm{pc}^{-1}$. To compute the mass of filament we employ this relation:
\begin{equation}
  M_\mathrm{fil} = \sum_{i=1} N(\rm H_2)_i \times  \mu \times m_H \times A.
\end{equation}
In this case, we obtain $M_\mathrm{fil} = 1.04  \mathrm{M}_\odot$, where N(\hh) is gas column density of the $i_\mathrm{th}$ pixel and A is the area of the pixel; $\mu=2.8$ and $m_\mathrm{H}$ are gas mean molecular weight per hydrogen molecule and hydrogen mass, respectively.
Therefore, we quantify the mass accretion rate along the filament, $\dot{M}_\mathrm{acc} = M_\mathrm{fil} \times \nabla V = 9.7 \times 10^{-7} \mathrm{M}_\odot \mathrm{yr}^{-1} $. 
This value is affected by some sources of uncertainty. We consider a 2\% error on the distance \citep{dzib2018}, a 12\% error on the calibration, and a 40\% error due to the assumption of the dust opacity index (\citealt{benedettini2018}, \citealt{roy2014}). The uncertainty on the mass estimation with all these three errors is a total of 42\%. With the 26\% error on the velocity gradient, we get a final relative error of 49\% for the value of the mass accretion rate. In addition, the inclination of the filament with respect to the plane of the sky is unknown, and it has an influence on the value of $\dot{M}_\mathrm{acc}$ by a factor of tan i (see, e.g., \citealt{chen2019}). The derived value of the accretion rate changes by  up to 70\% if the inclination varies between 30 and 60 degrees. Therefore, considering these factors, we assume that the derived $\dot{M}_\mathrm{acc}$ value is accurate within a factor of two.
This mass accretion rate is comparable to the value that \cite{pineda2020} found for a streamer of material connecting a protostar, Per-emb-2 (IRAS 03292+3039), to the surrounding cloud ($ \dot{M}_\mathrm{acc} = 10^{-6} \mathrm{M}_\odot \mathrm{yr}^{-1}$). The accretion rate that we calculate here is for a filament, not a streamer. However, this filament is diffuse and not very dense. Furthermore, the value we found is comparable to the mass infall rate of protostellar envelopes estimated in other young objects $ \sim 3 \times 10^{-6}$ \citep{evans2015}.

We detect another velocity gradient around the protostar in the west--east direction, which could be due to the rotation of the core. In this scenario the rotation axis would lay in the north--south direction (PA = 16$^{\circ}$; see last paragraph of the section), close to the direction of the detected bipolar outflows found by \cite{bjerkeli}, which are shown with the blue and red arrows in the bottom panel of Fig. \ref{velocity} (PA = 35$^{\circ}$). \cite{redaelli2019} also showed that this outflow direction is almost parallel to the mean magnetic field, which has the direction of PA=45$\pm7 ^{\circ}$. The black vectors in the bottom panel of Fig. \ref{velocity} represent the polarization angles, tilted by 90 degrees to trace the magnetic field direction. The rotation axis and magnetic field lines are moderately well aligned (with an offset of 29$^{\circ}$). According to the magnetohydrodynamic (MHD) collapse models, magnetic braking should be effective in this cloud (\citealt{joos2012}, \citealt{li2013}, \citealt{krumholz2013}, \citealt{seifried2015}), which is consistent with the absence of a resolved Keplerian disk (down to 30 AU, \citealt{yen2017}).
\cite{okoda2018A} observed IRAS15398-3359 with better resolution (0.2'' angular resolution). A disk of no more than 30 au in size was detected in their analysis with a mass between 0.006 and 0.001 $\mathrm{M}_\odot$.  This is a very small disk that is consistent with magnetic braking.
Magnetic braking is in fact an efficient way to remove angular momentum from infalling and rotating material, suppressing envelope fragmentation and the formation of large disks \citep{li2014}.

The velocity gradient over the protostar shows gas motions likely consistent with the rotation of the core. To analyze the gas motions further, we employ a two-dimensional representation of the velocity gradients. Figure \ref{gradient_velocity} shows the centroid velocity of \dco (3-2), overlaid with the gradient velocity arrows. The arrow length represents the relative vector magnitude of the gradient and the direction of the arrows point in the direction of increasing velocity. The method used here followed the analysis of \cite{goodman1993}, and developed by \cite{caselli2002a}, to find the velocity gradient based on a linear fit between offset declination and the velocity. By assuming that  the core rotates as a solid body, $V_\mathrm{LSR}$ would only depend on the coordinates in the sky and not on the distance along the line of sight. In this approximation, the centroid velocity of the line is a linear function of the offset on the plane of the sky
\begin{equation}
  V_\mathrm{LSR} = V_0 + a \Delta \alpha + b \Delta \delta  
,\end{equation}
where $\Delta \alpha$ and $\Delta \delta$ represent offsets in right ascension and declination, and $V_0$ is the systemic velocity of the cloud with respect to the local standard of rest. The coefficients $a$ and $b$, together with $V_0$, can be obtained by least-squares fitting. The velocity gradient magnitude is then \citep{goodman1993}
\begin{equation}
   \nabla V_\mathrm{LSR} = \sqrt{(a^2 + b^2)} 
,\end{equation}
and its direction toward increasing velocity $\Theta_{\nabla V}$ is given by

\begin{equation}
   \Theta_{\nabla V} = \tan^{-1}{ \frac{a}{b}} 
.\end{equation}
We use this method and obtain the velocity gradient and the position angle with their uncertainties. The number of pixels used to carry out the fit is 9, which is appropriate for single-dish data that is Nyquist sampled \citep{caselli2002a}.

We only consider values with S/N > 3 in the final result for each velocity gradient value; the mean signal-to-noise ratio around the protostar position is 7.
The mean velocity gradient magnitude distribution around the core peaks at $5.1 \pm 0.7$ \kms pc$^\mathrm{-1}$ and has a mean position angle of $(106\pm 7)^{\circ} $ (counterclockwise from the north toward the east), which is shown as a yellow arrow in Fig. \ref{velocity}. We can see that this mean velocity gradient is consistent at the 3$\times \sigma$ level to be perpendicular to the direction of the bipolar outflow found by \cite{bjerkeli} (PA=35$^{\circ}$).  We assume that the total gradient is the rotational direction of the core.
This method is implemented in a Python code which is available via  open access on GitHub.\footnotetext[5]{\url{https://github.com/jpinedaf/velocity_tools}}\footnotemark[5]


\subsection{ \co (2-1) line}

Figure \ref{C18O_V} represents the centroid velocity map obtained fitting the observed \co (2-1) spectra. The left panel of Fig. \ref{C18O_V} represents the brightest components of the Gaussian fitting, and as we discussed in Sect. 4.2, in some locations of the cloud \co (2-1) gas reveals two velocity components. The right panel of Fig. \ref{C18O_V} shows the second component, the one with lower intensity. Depending on the location on the map, this less bright component has red or blue velocities compared to the main component. The right panel of Fig. \ref{C18O_V} shows that by moving from north to south the faint component is going to the red side of the brightest components. The red velocities appear mostly on the south of the filament and the west side of the protostar position. This is similar to our discussion in Sect. 4.2, where there are three components along the line of sight: the bright one associated with the core and two fainter ones, one in the north and one in the south of the filament.

A small velocity gradient can also be seen in the brightest component of the \co (2-1) emission around the protostar in the east--west direction, in agreement with what we discuss in Sect. 4.3 for the \dco (3-2) line, which is likely associated with the rotation of the core. For positions where N(\hh) $> 2 \times 10^{22}$ \cm, the mean value of the centroid velocity is $\langle V_\mathrm{lsr} \rangle = 5.22$ \kms with an uncertainty 0.04 \kms. The $V_\mathrm{lsr}$ value at the west side of the core is $5.169\pm 0.005$ \kms and increases toward the east side of the core up to the value $5.290\pm 0.003$  \kms. These two values are calculated at the edge of the core, where the N(\hh) is equal to $2 \times 10^{22}$ \cm (the inner contour in Fig. \ref{C18O_V}).

\citet{frau} proposed for the Pipe nebula that the sharp changes in the magnetic field is produced by shocks between two clouds and, in comparison to the non-shocked gas, the column density and magnetic field strength double. \citet{redaelli2019} observed a sharp change in the magnetic field toward dust extension in the northwest direction of this cloud with respect to the core magnetic field. We predict that this is produced by merging two distinct clouds, two components of the \co (2-1) line. These polarization vectors, which show different directions concerning the vectors in the core position, could result from the cloud collision.

\subsection{Comparison between the \co (2-1) and \dco (3-2) kinematic}

\begin{figure}
   \centering
   \includegraphics[width=\hsize]{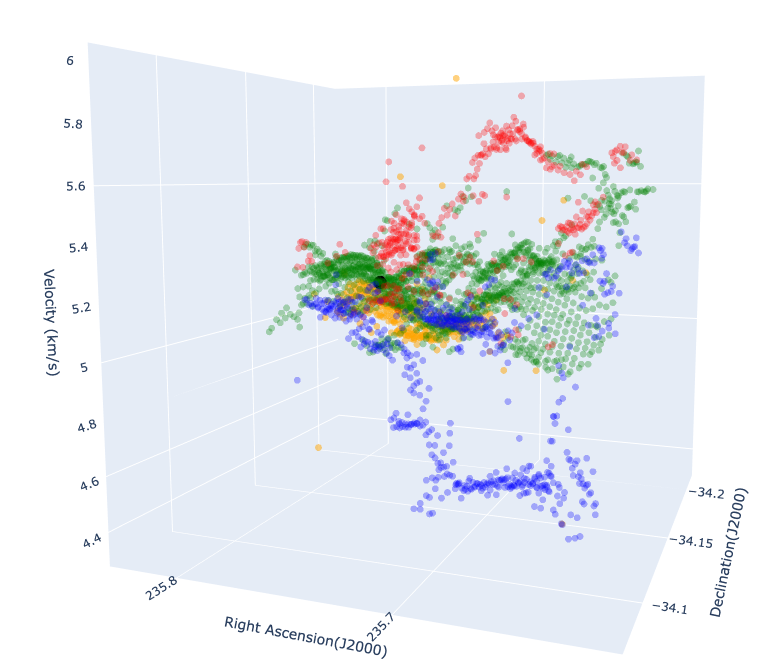}
      \caption{PPV image of \co (2-1) and \dco (3-2) gas. Each data point denotes the location and centroid velocity of a Gaussian component and each color refers to a different Gaussian fit. The  the centroid velocity is shown for the \dco (3-2) line (in orange), and the green data points represent the brightest component of the \co (2-1) line. The red and blue   data points indicate the two lower intensity velocity components. The black circle represents the position of the protostar.
              }
         \label{ppv_plot}
   \end{figure}

 The result of the fitting procedure is shown in Fig. \ref{ppv_plot}, where we present an image depicting a 3D position-position-velocity (PPV) diagram highlighting the distributions of \dco (3-2) and \co (2-1) gas throughout the cloud.
Each data point illustrates the location and centroid velocity of an independent Gaussian component. The color of each data point relates to each spectral component (discussed in Sect. 4.2).
Orange refers to \dco (3-2) emitting gas, and the others refer to \co (2-1) emitting gas. The velocity structure of the \co (2-1) emission is quite complex. Four velocity components are displayed in total. Overall, for \co (2-1) we used only two-component fitting, but here we display it in different categories. The blue data correspond to the brightest components of the \co (2-1) gas, and the red and cyan points instead are related to the secondary fainter component, when it is located at lower and higher velocities, respectively, with respect to the brightest one.

We observe a systematic difference between the centroid velocity of \dco (3-2) and the main component of \co (2-1), suggesting that they are not tracing exactly the same gas. We speculate that this is related to the fact that one is an ion and the other is a neutral species, and they behave differently concerning the magnetic fields, or it could be because of the difference between their gas densities. It is important to note that if ions and neutrals behave differently at the same density, this indicates a violation of the flux-freezing assumption. We   discuss   this point in more detail in Sect. 5.
 
\begin{figure}
   \centering
   \includegraphics[width=\hsize]{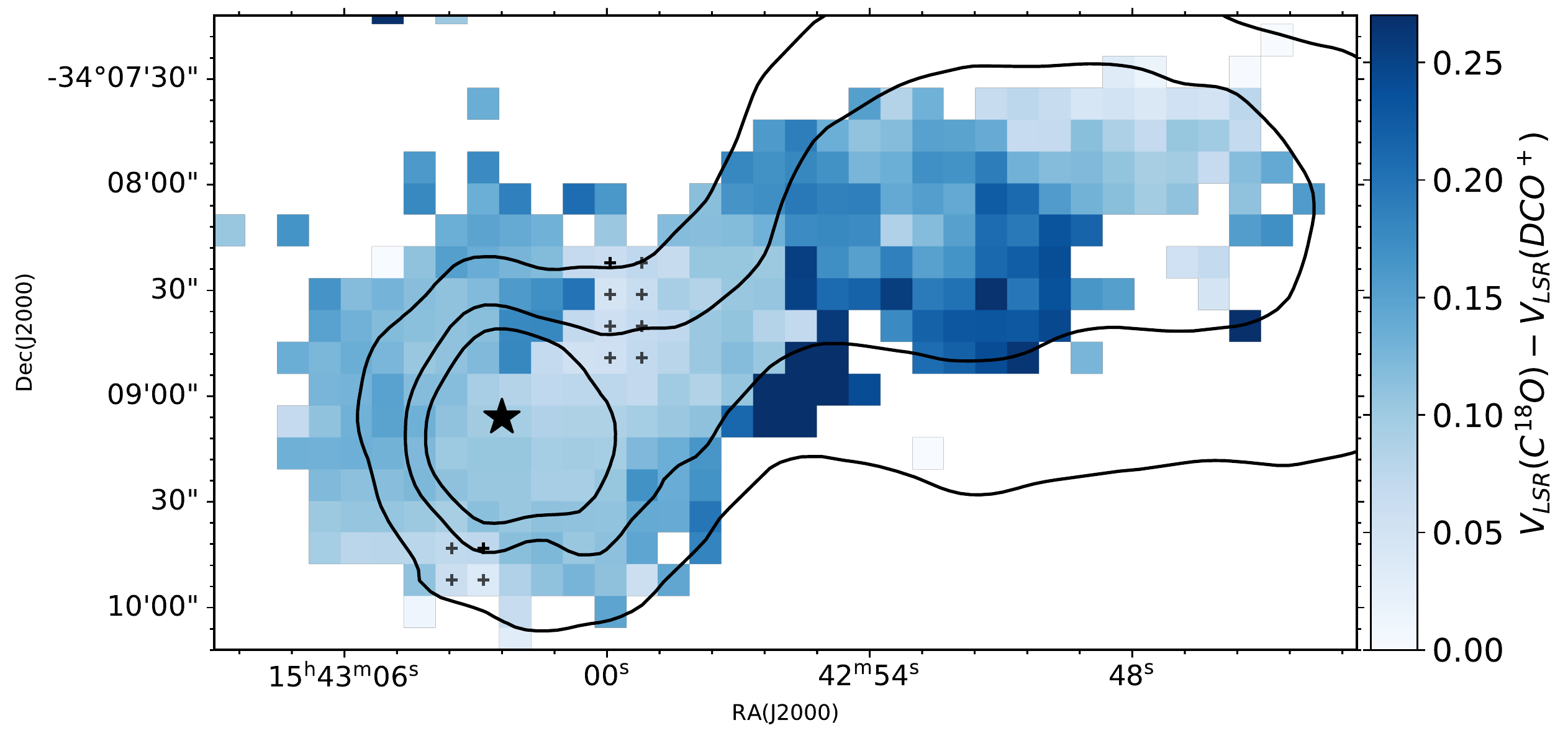}
      \caption{ Map of velocity shift between \co (2-1) and \dco (3-2). Overlaid in black contours is the \hh column density (levels:[1.0,1.5,2.0] $10^{22} $ \cm). The black star represents the position of the protostar and the black plus signs are positions of the lowest velocity shifts, which are used for the mean value.
              }
         \label{velocity_shift}
   \end{figure}

Figure \ref{ppv_plot} shows that the main velocity component of the \co (2-1) data and the \dco (3-2) line present a systematic velocity shift. The \co (2-1) data   always appear at a higher velocity than the \dco (3-2) spectra. In order to investigate this point further, we show in Fig. \ref{velocity_shift} the velocity difference $V_\mathrm{lsr}$(\co) - $V_\mathrm{lsr}$(\dco) with the \hh column density overlaid on top. In addition, we did not find any correlation between the velocity shift and the \hh column density map. In this figure the shift between \co (2-1) and \dco (3-2) is clearly visible. We report a mean velocity shift of 0.13 \kms across the whole source. 
The largest velocity shift values are found on the west side of the protostellar core, where the infalling material from the envelope reaches the core, under the assumption that the small velocity gradient seen along the filament represents an accretion flow. 
 On the southeast and northwest sides of the core there is a very low-velocity shift, equal to $\sim$ 0.06 \kms and these values are increasing toward the protostar position up to  0.10 \kms. 
 To calculate these low-velocity shifts we use the pixels indicated with a black plus sign in Fig. \ref{velocity_shift}.

\section{Discussion}

Star-forming regions can be more completely understood by analyzing the distribution of different molecular species in the velocity. We compare the velocity shift between these two tracers in the cloud. The velocity shift between a neutral and an ionized species was observed in the past. For instance, \cite{henshaw2013} studied the large-scale velocity field throughout the cloud in G035.39-00.33, and found a velocity shift between the two tracers of N$_2\mathrm{H}^+$(1-0) and  \co (1-0), in agreement with a model of collision between filaments that is still ongoing. It follows that the velocity structure of the core does not have intrinsic properties but is a product of large-scale motions on filamentary scales. They proposed that the velocity difference in the cloud occurs because of filament merging, implying that higher velocity filaments are interacting with a lower velocity, less massive filament, increasing the density of an intermediate velocity filament \citep{barnes2018}.

Another scenario is that the velocity shift between the \co (2-1) and \dco (3-2) reveals relative motions between the dense gas, traced by \dco (3-2), and the surrounding less dense envelope, traced by \co (2-1). \dco (3-2) is simply tracing higher densities, which may not necessarily have the same velocities as the gas traced by \co (2-1), especially because the \co (2-1) emission is much more extended, so the kinematics derived from the line profile is affected by lower density material not seen in \dco (3-2). According to \citealt{zhang2017}, velocity shifts between high-density and low-density tracers within a cloud are indicative of gas expanding and contracting. It is based on the assumption that the higher critical density molecules trace the dense gas closer to the inner of a core, while lower critical density molecules trace the more extended gas in the outer envelope.

Magnetic braking might have had a great impact in this cloud. We determined that the rotation axis of the core and magnetic field lines are almost aligned. According to the magnetohydrodynamic (MHD) collapse models, magnetic braking should be effective in this cloud, which is in agreement with the absence of a resolved Keplerian disk. Magnetic braking is an effective way to remove angular momentum from infalling and rotating material, suppressing envelope fragmentation and the formation of large disks.

%


\section{ Conclusions}

We studied the region around the young low-mass Class 0 source  IRAS15398 using APEX. The kinematic analysis performed with the \co (2-1) emission line as a low-density material tracer (extended gas) and with the \dco (3-2) line as a tracer of dense gas closer to the protostar. The measured kinematics parameters revealed several properties by performing Gaussian fitting, using the \textsc{pyspeckit} package. Our main conclusions can be summarized as follows:

\begin{description}
  \item[$\bullet$ ]  From the spectral line profiles we conclude that the two velocity components of \co (2-1) in the west side of the region have merged together toward the position of the protostar.
  \co (2-1) is a lower-density tracer than \dco (3-2); the fainter velocity component seen only in the former tracer is likely low-density filamentary material associated with the cloud. The core envelope appears to be located in correspondence of the merger of these filamentary structures.

  \item[$\bullet$ ] We see a velocity gradient along the filament in the \dco (3-2) gas. Therefore, we measured the ongoing accretion material toward the protostar core in this gas, $\dot{M}_\mathrm{acc} = 9.7 \times 10^{-7} \mathrm{M}_\odot \mathrm{yr}^{-1}$, where the accretion rate is expected to be accurate within a factor of 2.

  \item[$\bullet$ ] The mean velocity gradient is roughly 5.1 \kms pc$^{-1}$ measured in \dco (3-2) around the protostar core, which is linked to the rotation of the core. This velocity gradient at the position of the protostar is in the east--west direction, oriented approximately perpendicular to the bipolar outflow previously found.

  \item[$\bullet$ ] Line widths of \dco (3-2) increase toward the position of the protostar, probably due to protostellar feedback.

  \item[$\bullet$] We observed a velocity shift between neutral and ionized species. A higher velocity is always present in the \co (2-1) data compared to the \dco (3-2) data. The mean velocity difference, $V_\mathrm{lsr}$(\co) - $V_\mathrm{lsr}$(\dco), is equal to 0.13 \kms across the full filament. This is consistent with a model of collision between filaments that is still ongoing. The velocity shift between the \co (2-1) and \dco (3-2) illustrates the relative motion of the dense gas, traced by \dco (3-2), and the surrounding less dense envelope, traced by \co (2-1).
  
\end{description}

Further observational investigations are needed to determine in more detail  the connections within the kinematics and magnetic field in this source.

\begin{acknowledgements}
      Elena Redaelli acknowledges the support from the Minerva Fast Track Program of the Max Planck Society. The authors would like to thank Jaime Pineda Fornerod for his support and discussion about the code to calculate the velocity gradient. This research has made use of data from the Herschel Gould Belt survey (HGBS) project (http://gouldbelt-herschel.cea.fr). The HGBS is a Herschel Key Programme jointly carried out by SPIRE Specialist Astronomy Group 3 (SAG 3), scientists of several institutes in the PACS Consortium (CEA Saclay, INAF-IFSI Rome, and INAF-Arcetri, KU Leuven, MPIA Heidelberg), and scientists of the Herschel Science Center (HSC)\citep{andre2010A}.
\end{acknowledgements}

%
%

\bibliographystyle{aa} 
\bibliography{aanda} 
%

%
%

\begin{appendix} 
\section{Positions of spectral grid }
\begin{figure}
   \centering
   \includegraphics[width=\hsize]{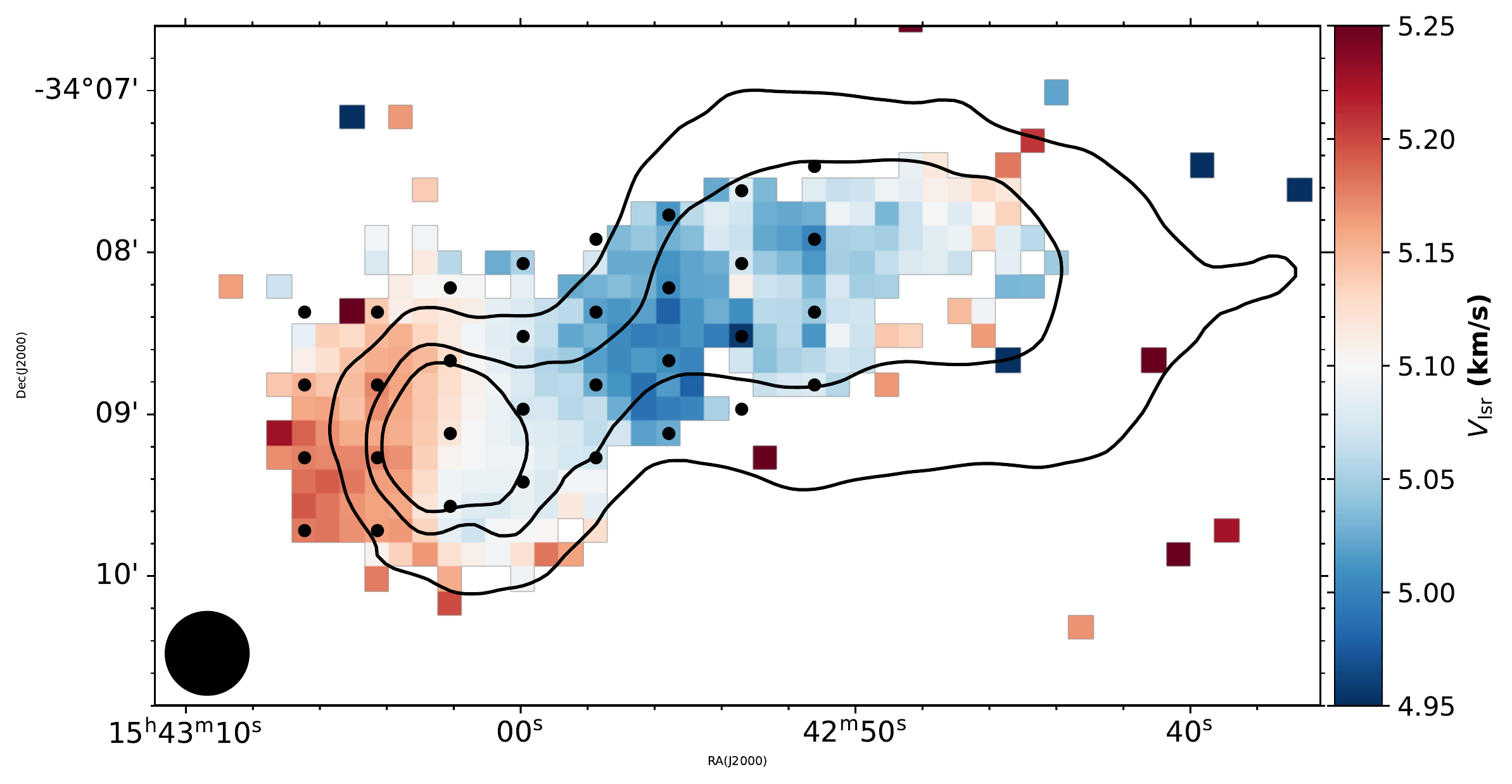}
      \caption{ Map of centroid velocity od \dco (3-2). The black dots showing the position of each spectrum. Overlaid in black contours is the \hh column density (levels:[1.0,1.5,2.0] $10^{22} $ \cm). 
              }
         \label{positions}
   \end{figure}

Figure \ref{positions} shows  the position of each spectra in Fig. \ref{channel}. There are 40 black dots that show the exact location of each spectrum at an 18 arcsec interval from each other.

\end{appendix}

\end{document}